\newif\ifabstract
\abstracttrue
\abstractfalse 
\newif\iffull
\ifabstract \fullfalse \else \fulltrue \fi

\documentclass[11pt]{article}
\usepackage{amsfonts}
\usepackage{amssymb}
\usepackage{amstext}
\usepackage{amsmath}
\usepackage{amsthm}
\usepackage{algorithm}
\usepackage{algorithmic}
\usepackage{tikz}
\usepackage{multicol}
\usetikzlibrary{calc}
\usetikzlibrary{arrows,shapes}
\usepackage{dot2texi}
\usepackage{siunitx}
\usepackage{booktabs}
\usepackage{pdflscape}
\usepackage{threeparttable}
\usepackage{longtable}
\usepackage{multirow}
\usepackage{blindtext}
\usepackage[inline]{enumitem}
\usepackage{xcolor}

\usepackage{xspace}
\usepackage{theorem}
\usepackage{graphicx}
\usepackage{url}
\usepackage{graphics}
\usepackage{colordvi}
\usepackage{colordvi}
\usepackage{lscape}
\usepackage{natbib}
\usepackage[english]{babel}
\usepackage{float}
\usepackage{caption}
\usepackage{subcaption}
\usepackage{geometry}
\usepackage{amsmath}

\advance \oddsidemargin by -0.0001in
\newcommand{\myparskip}{3pt}
\parskip \myparskip

{\hspace*{\fill}$\Box$\par\vspace{4mm}}

\begin{document}

\title{Perturbation Privacy for Sensitive Locations in Transit Data Publication: A Case Study of Montreal Trajet Surveys \footnote{\textbf{Transportation Research Records of Transportation Research Board}}}
\author{Godwin Badu-Marfo  \thanks{TRIP Lab, Concordia University, Email: {godwin.badu-marfo@mail.concordia.ca}} \and Bilal Farooq \thanks{Laboratory of Innovations in Transportation (LITrans), Ryerson University, Email: {bilal.farooq@ryerson.ca}} \and Zachary Patterson \thanks{TRIP Lab, Concordia University, Email: {zachary.patterson@mail.concordia.ca}}}

\begin{titlepage}
\maketitle

\thispagestyle{empty}

\begin{abstract}
Smartphone based travel data collection has become an important tool for the analysis of transportation systems. Interest in sharing travel survey data has gained popularity in recent years as ``Open Data Initiatives" by governments seek to allow the public to use these data, and hopefully be able to contribute their findings and analysis to the public sphere. The public release of such precise information, particularly location data such as place of residence, opens the risk of privacy violation. At the same time, in order for such data to be useful, as much spatial resolution as possible is desirable for utility in transportation applications and travel demand modeling. This paper evaluates geographic random perturbation methods (i.e. Geo-indistinguishability and the Donut geomask) in protecting the privacy of respondents whose residential location may be published. We measure the performance of location privacy methods, preservation of utility and randomness in the distribution of perturbation distances with varying parameters. It is found that both methods produce distributions of spatial perturbations that conform closely to common probability distributions and as a result, that the original locations can be inferred with little information and a high degree of precision. It is also found that while Achieved K-estimate anonymity increases linearly with desired anonymity for the Donut geomask,  Geo-Indistinguishability is highly dependent upon its privacy budget factor ($\epsilon$) and is not very effective at assuring desired Achieved K-estimate anonymity. 
\end{abstract}

\end{titlepage}

\section{Introduction}
Transportation demand modeling helps governments and researchers to better understand human mobility in the delivery of an efficient, intelligent and secure transport system and is highly dependent on quality travel demand data. Traditionally, travel demand surveys (Origin-Destination, regional, household, etc.) actively engage respondents in the collection of travel and personal information. In such contexts, collected data is available to relatively few institutions and individuals who are typically employed and under contract not to use the data for any purposes apart from those strictly related to their responsibilities (e.g. transportation planning). These institutions have also been responsible for ensuring the protection of privacy when any of the data is provided to other institutions, or the public, i.e. when the data are ``published.''

Nowadays, there is a proliferation of pervasive devices and technologies (e.g. smartphones, tablets, wearable devices, etc.), with location-sensing capabilities. Travel Surveys are now delivered on these technologies that can collect personal and sensitive information (i.e. biographical data, credit card information, location traces, etc.) passively even without respondents being aware. The mobility of a respondent is typically recorded as trajectories and processed by location based services (e.g. Google\citep{google2018online}, Uber\citep{uber2018online}), location-aware applications (e.g. Waze\citep{waze2018online}) or dedicated travel survey apps \cite{patterson2017itinerumtrip}). Whilst governments, public and private transport researchers exploit the potential of passive large-scale transport data to understand mobility patterns and travel demand, the threat of personal information disclosure cannot be overlooked. 

As witnessed in recent years, numerous high-profile privacy breaches have taken place. There was, for example, an enormous public outcry around the privacy controversy related to Facebook and Cambridge Analytica in 2018 \citep{olivia2018facebook}. Another example was provided by Anthony Tockar \citep{neustar2018riding}, a summer intern at Neustar (an information-analytics company) who showed how to extract the exact location and time that celebrities used cabs in New York City based on publicly available New York City Taxi and Limousine Commission (TLC) data. By joining the two data sets, Tockar was even able to find the cash tips paid by celebrities \citep{neustar2018riding} to their drivers. These examples have given rise to an interest in data privacy violations and the need for ``data agents" \cite{xu2014information} (data collectors and analyst individuals or organizations) to protect personally identifiable information. 

At the same time, governments have been eager to adopt ``Open Data Initiatives" that make data available for free reuse and republishing to everyone, without restrictions related to copyright or patents \citep{gurstein2011open, kitchin2014data}. Open data agreements between governments, transport operators and travel application developers have been witnessed in the sharing of information for improving transportation service delivery. The City of Toronto in 2017 entered into an agreement with Waze \citep{toronto2017waze} to share and use its real-time traffic and road conditions data, to improve service delivery and navigation in the city of Toronto. Uber also launched ``Uber Movement" \citep{uber2018movement} a platform that shares travel information with cities and transport planners with the aim of helping them make informed decisions in the design of transport infrastructure.  

Since there are no controls on who can access Open Data, any sensitive data provided by people on whom data is collected i.e. ``Data Owners'' \cite{xu2014information}, such as respondent identity needs to be protected to prevent privacy breaches by untrusted users with malevolent intentions, or ``adversaries." Geographic points of interest (POIs) can be extracted from trip data and inferences can be made on  characteristics (i.e. semantic data) such as religious affiliation, health conditions and political interests of respondents. In line with this, the disclosure of sensitive location information poses a risk and could violate a respondent`s confidentiality if known to an adversary or untrusted party. Thus as governments adopt ``Open Data" policies, it is important for the location privacy (or geoprivacy) of a subject to be protected to ensure the identity of an individual is not disclosed through location information.

Location protection mechanisms \citep{krumm2009survey} such as spatial cloaking, aggregation and random perturbation are used to protect--what we refer to as--the Personally Identifiable Location Information (PILI) of a subject \cite{badu-marfo2018perspectives}. Random perturbation techniques endeavor to deliver better privacy, while maintaining spatial fidelity of data to maximize the utility of anonymized spatial data \citep{kounadi2015spatial, allshouse2010geomasking}, while protecting privacy. Random perturbation methods are used to add noise that displaces/masks point locations in a random distance and direction. A popular random perturbation method, geomasking, is used for preserving location privacy by creating a circular buffer at a specified distance around the location to be masked, from which the perturbed location is selected. Geomasking is the most common method of perturbing an individual`s location for privacy protection\cite{allshouse2010geomasking, hampton2010mapping}. However, the quantity of displacement applied to a location for masking can at once reduce the utility of the data and, if displacement is small, provide little privacy protection. Among the various random perturbation techniques, two have received the greatest amount of attention and are the most commonly used in practice.

The first, the ``Donut Geomask'', which is an implementation of a k-anonymity location privacy protection mechanism (LPPM) \citep{sweeney2002k} and achieves privacy protection by using the underlying neighborhood population density of a point location to determine the obfuscation distance. This geomask technique has been used extensively in the protection of patient health information \citep{hampton2010mapping,allshouse2010geomasking} and crime data \citep{kounadi2015spatial}, both of which require high spatial resolution in order to anonymize data.

The second, Geo-indistinguishability (Geo-I) \citep{andres2013geo}, is an implementation of differential privacy for location data. It guarantees a respondent`s location is protected within a specified protection distance with a level of added noise that decreases with the distance, at a rate depending on the desired level of privacy. In other words, the original location is highly indistinguishable from locations that are close to it, and gradually becomes more distinguishable from locations that are farther away \citep{chatzikokolakis2015location}. This is intended to maintain anonymity, while at the same time maintaining the utility of the underlying data.

In this paper, we have three main contributions. First, we apply the most common geographic anonymization techniques to the case of residential location of respondents in a large-scale smartphone travel survey, MTL Trajet \citep{mtltrajet2018web}. Second, we evaluate both techniques with respect to their ability to provide location privacy while maintaining utility of the data. Third, we analyze the distribution of the perturbation distances for their degree of randomness to evaluate the degree to which it would be possible to infer original location by an adversary with prior knowledge of only the distribution of disturbances.   	

\section{Problem Statement}
The objective of this paper is to protect respondent residential locations collected in a travel survey before data are published. We compare and evaluate the two most commonly used random perturbation techniques (the Donut Geomask and Geo-I) to measure and their efficiency of privacy protection and their effectiveness of data utility. We aim at evaluating the degree of protection offered by both techniques by studying the probability distribution of the achieved perturbation distances. 

To achieve this objective, we consider respondents who took part in the MTL Trajet smartphone survey of 2016. MTL Trajet respondents were asked to report their home location(latitude and longitude) as part of the survey. We treat home locations as sensitive, independent points that need to be protected to prevent violation by an adversary who has access to the published travel data. (Note that these residential location data was never published by the City of Montreal.) 

\section{Literature Review}
Extensive literature exists on the geographic perturbation of location addresses for privacy protection. This research has been undertaken by numerous disciplines (e.g. computer science, geography, transportation, etc.) that are engaged in dealing with  personal identifiable location information (PILI). We discuss a few examples of this in the following section.

Zhang et al. \citep{zhang2017location} developed a geomasking technique referred to as location swapping. Their technique replaces an original location with a masked location that is selected from all possible locations with similar geographic characteristics within a specified neighborhood. Their technique provided greater anonymity than other random methods by achieving higher k values. (K is the population around the sensitive location that could be associated with equal probability to the perturbed location. If K, were for example 10, then the perturbed location could be equally attributable to 10 different households.)

Allhouse et. al. \citep{allshouse2010geomasking}, used the Donut method to provide privacy for sensitive health data using household data for Orange County, North Carolina. The authors determined the actual k-anonymity (the number of households that could be associated with perturbed points) by revealing household locations contained in the county database. They achieved an approximate privacy standard for the households at 99.5\% (i.e., 99.5\% of perturbed points represented k-anonymities of above a desired threshold).

Abul et al. \citep{abul2008never}, proposed a technique that creates cylinders within which users move such that at every instant of time, there exists at least k users walking a given distance from others.

Ma et al., \citep{ma2014nearby} implemented Geo-I in protecting the privacy on nearby friend-request location-based services (LBS for short) from stalkers. The authors combined the location approximation technique and the homomorphic cryptography to achieve formal privacy guarantees for LBS users, and achieved a satisfactory quality of the reported location to be used by the LBS. That is, the query results were relevant to the original location of the user, even though only perturbed data was provided to the LBS.

Finally, Chatzikokolakis et al. \citep{chatzikokolakis2015location}, protected the privacy of LBS users using the principle of Geo-I. Using the foundations of Differential-Privacy, their work protected exact user location, while providing  sufficiently accurate location information to allow satisfactory results to be provided by the LBS.

\section{Background on Anonymization Techniques Considered} \label{sec:techniques}
As described above, the two most widely used classes of anonymization techniques are k-Anonymity and Differential-Privacy. This section describes them in greater detail.

\subsection{K-Anonymity}
K-Anonymity is the most widely used class of privacy protection technique for location-based systems existing in literature. The notion of k-anonymity was introduced by Sweeney in 2002 \citep{sweeney2002k}. Many implementations of k-anonymity aim at protecting a subject`s identity, requiring that an adversary cannot identify an individual record, among a set of k indistinguishable subjects (i.e. any query result in no less than k observations). k-Anonymity has been used in protecting location (l-diversity), that requires that a set of k-points are spatially indistinguishable. This technique of achieving location privacy using k-anonymity can be implemented through the use of dummy locations \citep{cheng2006preserving, bamba2008supporting}, where k-1 dummy points are generated and returned as a location-based query result \citep{bamba2008supporting}. Another implementation to achieve k-anonymity in location privacy is through the use of spatial cloaking \citep{gruteser2003anonymous}. This approach creates a cloaking region around the real location point with k other points. The cloaking region is then returned as the result of a location-based query and protects the original location by making it indistinguishable among a set of k points in its cloaked region \citep{gruteser2003anonymous}. The first technique we compare examines the random perturbation of sensitive locations using a new adaptive geomasking technique, referred to as the Donut method which has the property of k-anonymity. 

\subsubsection{Donut Geomask} \label{sec:donut}
\begin{figure}[!h]
	\begin{subfigure}{.5\textwidth}
		\centering
		\includegraphics[width=.8\linewidth]{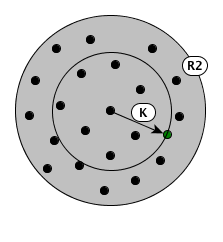}
		\caption{Estimated k-Anonymity}
		\label{fig:actualk}
	\end{subfigure}%
	\begin{subfigure}{.5\textwidth}
		\centering
		\includegraphics[width=.8\linewidth]{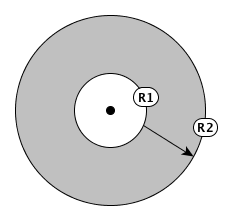}
		\caption{Donut geomask}
		\label{fig:Donut}
	\end{subfigure}
	\centering
	\begin{subfigure}{.5\textwidth}
		\centering
		\includegraphics[width=.8\linewidth]{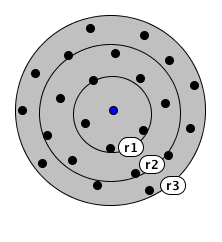}
		\caption{Varying protection distance, r}
		\label{fig:varying_dist}
	\end{subfigure}
	\caption{Examples of:  (a) Calculating the estimated k-anonymity of a location. (b) Generating protection distance by the Donut approach, and (c) Varying protection distances with Max K.}
	\label{fig:fig}
\end{figure}

The Donut method is a geomask technique that protects the privacy of locations by transposing real locations to random displacements within an inner circle radius (i.e. a minimum limit of perturbation) and an outer circle radius that is the maximum limit of the perturbation distance. As illustrated in Figure \ref{fig:Donut}, \textit{R1} is the inner circle representing the minimum displacement from the original location. This method prevents negligible displacements that are close to the original location. On the other hand, \textit{R2} is the outer circle which sets the maximum distance to of random displacement \citep{allshouse2010geomasking}. Random displacements of perturbed points are inversely proportional to the underlying population density, and this guarantees privacy protection of point locations while minimizing spatial error \citep{zhang2017location,allshouse2010geomasking}. As an example, while point locations in urban high-density communities will only need to be perturbed small  distances, locations in low-density (e.g. rural) areas will need to be perturbed larger distances to achieve the desired level of privacy. The method provides a robust privacy guarantee as maximum and minimum thresholds of displacements are used to prevent negligible or outlier perturbation distances.

\subsection{Differential Privacy}
Differential privacy \citep{dwork2008differential} has gained popularity as a new privacy model for protecting an individual without disclosing the data of a subject when the subject participates in a database, and similar disclosure occurs with same probability when the subject does not participate in the database. This ensures that the removal or addition of any record about an individual in a database does not modify the results of a query. Intuitively, the concept of differential privacy requires that the distribution of the characteristics of two datasets (i.e. the original dataset, D and adjacent dataset, D`) differing by only one observation should not be noticeable. This is explained by the notion that an addition or removal of a single record in an adjacent dataset does not significantly affect the outcome of a query to the two datasets. 

It can be illustrated by a scenario where the probability of a query returning a value v when applied to a database D is similar when compared to the probability of reporting the same value to an adjacent database D`, differing by only one observation. The amount of difference between D and D` is parameterized as epsilon ($\epsilon$), or the ``privacy budget." In order to achieve differential privacy,  a controlled random noise is drawn from a Laplace distribution and added to a query output. Differential privacy has been applied in the context of location privacy, as observed in \citep{ho2011differential}. There a differentially private region quadtree is used for both de-noising the spatial domain and identifying the likely geographic regions containing the sensitive locations. The quadtree spatial decomposition enables one to obtain a localized, reduced sensitivity to achieve the differential privacy goal and accurate outputs. The most recent form of this technique is Geo-indistinguishability \citep{andres2013geo} and is the second technique we include in our comparison.

\subsubsection{Geo-indistinguishability} \label{sec:geoi}
Geo-indistinguishability (Geo-I) is a property similar to that of differential privacy \citep{dwork2008differential}. This privacy model is an implementation of differential privacy to address location privacy protection. Geo-I works with the notion that within a radius $r > 0$, a respondent is protected within r such that the level of privacy is proportional to the radius. This is illustrated by a basic scenario where a real location, \textit{li} is obfuscated by using some random noise to an approximate location that lies in radius r1 as shown in Figure \ref{fig:varying_dist}. At radius r1, a high level of privacy is achieved, making the real location indistinguishable among the nearest point locations (there are 3). At radius r2 and r3, the level of noise added to obfuscate \textit{li} decreases at a rate that is dependent on the desired level of privacy, ``epsilon." As an example, an adversary may be able to make a confident guess of the area where a respondent is located, but would not be able to predict the exact location of a respondent within the area \citep{chatzikokolakis2015location}. The random noise of perturbation for Geo-I can be implemented from a Laplacian distribution with respect to perturbation distance from original location. This approach is intended to ensure robustness with respect to the composition of attacks, as the level of privacy decreases in  a controlled way (linearly) \citep{chatzikokolakis2015location}. We implement an experimental simulation that achieves Geo-I by perturbing home locations of respondents in the MTL Trajet survey.

\section{Definitions}
In this section, we provide explicit definitions of technical concepts and key terms that will be used in the analysis of the paper. 

\textbf{Sensitive location} refers to any residential point locations represented in its Cartesian coordinates as latitudes and longitudes that needs to be protected to prevent the identification of a user.

\textbf{k-anonymity} refers to the population within a buffer region of the outer radius around the original point prior to displacement, from which a de-identified cluster case cannot be reversely identified. K is the population around a sensitive location that could be associated with equal probability to a perturbed location. If K, were for example 10, then the perturbed location could be equally attributable to 10 different households.

\textbf{Protection Radius} refers to a circular region around a sensitive location within which other location points existing should be made indistinguishable from the sensitive point.

\textbf{Location-privacy protection mechanism}. These are mechanisms that modify datasets to offer privacy guarantees by adding a level of noise to displace the sensitive location to distances away from their true location. Protected datasets are also referred to as \textit{geomasked datasets}.

\textbf{Adversary}. This is an agent seeking to re-identify true residential location of the user by inferring from sanitized dataset.

\section{Methodology} \label{methodology}
We assume a sensitive residential location, \textit{L}\textsubscript{r} of a respondent that needs to be protected by adding a random noise to displace the original location to a new location. We refer to the ``noised'' location as the "Perturbed Location" of the respondent. The distance and direction to which a sensitive location is displaced to guarantee protection is implemented by a location-privacy protection mechanism (LPPM). In this paper, we employ two LPPMs namely GeoI and Donut. 

A protection radius, r, is a required parameter for perturbation by both mechanisms and this sets the minimum distance to which a sensitive location will remain indistinguishable among a set of other locations nearby. As illustrated in Figure \ref{fig:Donut}, the Donut geomask method defines its protection radius as \textit{R1}, which is the minimum distance a sensitive location is displaced to ensure indistinguishability among a set of k points, referred to as "k-Anonymity". For the GeoI, the protection radius is the defined circular region around a sensitive location to which other locations within the radius are made indistinguishable by adding a level of noise. We employ a numerical set including 100, 200, 300, 400, 500 as the protection radius (or Max K) for both mechanisms. E.g. 100 is the radius required (the Max K) to ensure indistinguishability with 99 other locations, 200 is the radius required for 199 other locations, etc.

For the Donut geomask, an outer radius \textit{R2} is also defined to be the maximum distance that a sensitive location can be displaced. This limits the extent of perturbation for the Donut method.  We experiment by varying predefined k-anonymity levels in calculating sets of outer radii for each point as shown in Figure \ref{fig:varying_dist}. This allows us to experiment with how different perturbation radius sizes affect the output of desired anonymity results (i.e. the Achieved K-estimate). We undertake the Donut perturbation method of selecting random distance and angles within \textit{R1} and \textit{R2} using a random number generator built into the perturbation algorithm \ref{sec:expSetUp}. The distortion of the perturbation is guided by the region boundary such that a new geomask point does not fall outside the region of the original location. Using the desired k-anonymity level and population density, the outer radius \textit{R2} is calculated and inner radius, \textit{R1}, is estimated as 10\% of \textit{R2} in this paper. The outer radius, R2 varies from point to point since it depends on population density. As an example, for low density regions points are displaced at farther distances than in high density areas. 

For the Geo-I technique, we maintain the sets of outer radii \textit{R2} for the varying k-anonymity levels as the protection distance within which perturbation should occur. We undertake this approach to examine how a changing width of the perturbation region will affect the results of the geomasked points. To determine the sensitivity of the privacy budget epsilon, we employ varying privacy budget values (0.10, 0.20, 0.30, 0.40, 0.50) that are repeated for each protection distance. This range is typical of what is used in the literature. 

We compare the protection levels achieved from both techniques using the Achieved K  and Average error distance metrics. We then calculate the Euclidean distances between the original and geomasked points and summarize them in histograms to which we fit the following probability distributions (normal, lognormal, gamma, exponential, Weibull). Since there is a trade-off between privacy protection and the utility (or usefulness) of the perturbed data, we also evaluate the utility of the perturbed points. To do this, we use average spatial error (defined below). 

\section{Evaluation Metrics}
In this section, we present a set of metrics that we use in evaluating the effectiveness of the perturbation techniques used to protect true respondent residential location. We discuss the evaluation metrics under three main indicators: distribution of perturbation, location privacy and data utility. We discuss each of the metrics, the steps involved in its execution and the desired output of measure below.

\subsection{Location Perturbation Distribution}
To understand the effectiveness of the perturbation techniques used in this paper, we study the randomness in displacement for each perturbed point by calculating the euclidean distances between the original points and the obfuscated points for each technique. We refer to this as ``perturbed distance." The perturbed distances are tested for randomness of distribution by fitting five continuous distributions (Normal, Lognormal, Gamma, Weibull and Exponential), to assess if they follow a known distribution pattern. We consider perturbed distances that follow particular distributions as lacking randomness and showing weak privacy since, as shown by Farooq et al. \citep{farooq2013simulation} an adversary could reversely identity a subject`s residential location knowing only the parameters of the distribution. Distances which do not fit any distribution are considered to be strong against re-identification by an adversary. The fitted log likelihood values of the different distributions are presented to compare which distributions have the best fit.

\subsection{Location privacy}
In order to ensure a reliable guarantee of protection of a respondent`s location, we measure the success of privacy achieved in obfuscating such locations using two approaches: the achieved K-estimate and geographic distance. The first metric, Achieved K-estimate is a commonly used measure of privacy protection performance \cite{hampton2010mapping}. It is inspired from k-Anonymity \cite{sweeney2002k} and evaluates the accuracy of location privacy by measuring the number of households among whom a specific de-identified subject cannot be reversely identified \cite{cassa2006context, allshouse2010geomasking}. This is illustrated by the population of households that can be counted within a circular region with its radius defined by the euclidean distance from original location to its obfuscated location as shown in Figure \ref{fig:actualk}. 

The estimated k-anonymity for a sensitive location is calculated as:

$$ k_{est, i} = \pi \times  {D_i^2} \times \Bigg(\frac{N_i}{A_i}\Bigg) $$

where \textit{D}\textsubscript{i} is the measure of Euclidean distance between the original household location and its perturbed location, \textit{N}\textsubscript{i} is the population of the neighborhood block and \textit{A}\textsubscript{i} is the area of the geographic block of the neighborhood. The estimated k-anonymity metric replaces the exact location of a subject with an anonymized spatial region that contains at least K-1 other subjects preventing an adversary from distinguishing a unique subject at a probability of 1/K \cite{ghinita2010reciprocal}. Achieving higher Achieved K-estimate values guarantees a higher degree of privacy protection. In our analysis, we compare the derived Achieved K-estimate for each perturbed location and evaluate which perturbation technique provides the highest location privacy protection across a range of protection radii and varying privacy budgets. We consider a minimum of ten households to be the smallest to ensure privacy protection.

In our second approach, we use the geographic distance metric which evaluates the guarantee of privacy protection by the extent of perturbed distance. In using the euclidean distance between the original and perturbed location, we assume a small geographic distance offers a low level of privacy with weak protection guarantees, whereas for a stronger privacy guarantee, requires higher geographic distances. 

\subsection{Data Utility}
Discussions on privacy protection generally assume a trade-off between privacy and the usefulness of data after perturbation. In other words, when a higher privacy is achieved, the potential usefulness of the data is degraded especially in a transportation planning context. Consider an example where our application interest is transit assignment for metro users, and that we have information on a respondent`s home location that we would like to protect/perturb. The greater the perturbation distance to the home location, the more likely the perturbation is to protect the respondent`s identity. At the same time, the greater the perturbation distance, the more likely it will be that the respondent would be assigned to the wrong metro access station and therefore anonymized assignment results would not correspond with the original results. As such, we assess the utility of anonymized data for the purpose of travel modeling by comparing the amount of spatial error introduced by the perturbation techniques. In our analysis, a minimum observed average spatial error defines a high utility of sanitized data.  We calculated the average spatial error as:

$$ ASE = \frac{1}{n} \times \sum_{i=1}^nD_{i,j} $$

where n is the total number of point locations, \textit{D}\textsubscript{ij} is the euclidean distance from original location, \textit{i}, to perturbed location, \textit{j}.

\section{Experimental Setup} \label{sec:expSetUp}
For this analysis, we used training datasets as discussed in the section below. We built on algorithms and source codes that had been developed for both perturbation methods. For Geo-I, we enhanced the differential privacy algorithm which is implemented by Chatzikokolakis et al. \citep{chatzikokolakis2015location} in the Location Guard browser extension \citep{bordenabe2014measuring}. Our enhancement provided capabilities for varying epsilon and choosing an attribute field to derive protection distances of perturbation. The algorithm was deployed on Quantum GIS 2.18 \citep{quantum2013development} running on Microsoft Windows 10 desktop computer.  On the other hand, we used and improved existing algorithms for the Donut Geomask, which is built by the Bayesian Maximum Entropy Lab of the University of North Carolina \citep{bmelab2018source}. Our modification updated the libraries for the algorithm to be deployed on the Esri ArcGIS 10.4 platform. The complete source code is available at \url{http://github.com/gbmarfo/geoperturbation}.

\section{Experimental Results and Analysis}
\subsection{Training Datasets}
We conducted analysis on both perturbation methods (i.e. Donut and Geo-I) using the MTL Trajet data of 2016. MTL Trajet 2016 was a large-scale smartphone travel survey conducted by the City of Montreal using the app MTL Trajet developed by the Concordia University TRIP Lab. The study took place in October and November 2016 \citep{patterson2017mtlisctsc}. The residential location data used in the analysis came from the questionnaire asked to respondents after installation of the app. Altogether, the home locations of 7,985 respondents were included in the analysis. Whereas the data sets contain trajectory information of trips by users that we could infer sensitive origin destinations, we focused on working with the user`s reported data that had been definitively labelled as place of residence. Notwithstanding, the techniques and algorithms are applicable to sensitive trip origins and destinations as well. 

\subsection{Analysis on Privacy Protection}
In measuring the amount of privacy protection achieved, we calculated the Achieved K-estimate for perturbed points over varying protection distances for values as shown in Figure \ref{fig:varying_dist}, and evaluated the population that made them indistinguishable if reversely identified by an adversary. This is illustrated in Figure \ref{fig:actualk}. We experimented with varying protection distances to measure the relationship between privacy and perturbation distances. Our analysis on achieved privacy protection is categorized and discussed as follows.

\subsubsection{Achieved K-estimate Measurement}
We observe for the Donut approach, that the estimate of Achieved K-estimate increases linearly with an increase in protection distance. Thus, higher Achieved K-estimate values are obtained when perturbation distances increase. A higher Achieved K-estimate value guarantees a stronger privacy as a larger indistinguishable population is created around the perturbed point \citep{zhang2017location} (see Section \ref{methodology}). 

\begin{figure}[h]
	\centering
	\includegraphics[width=1.0\textwidth]{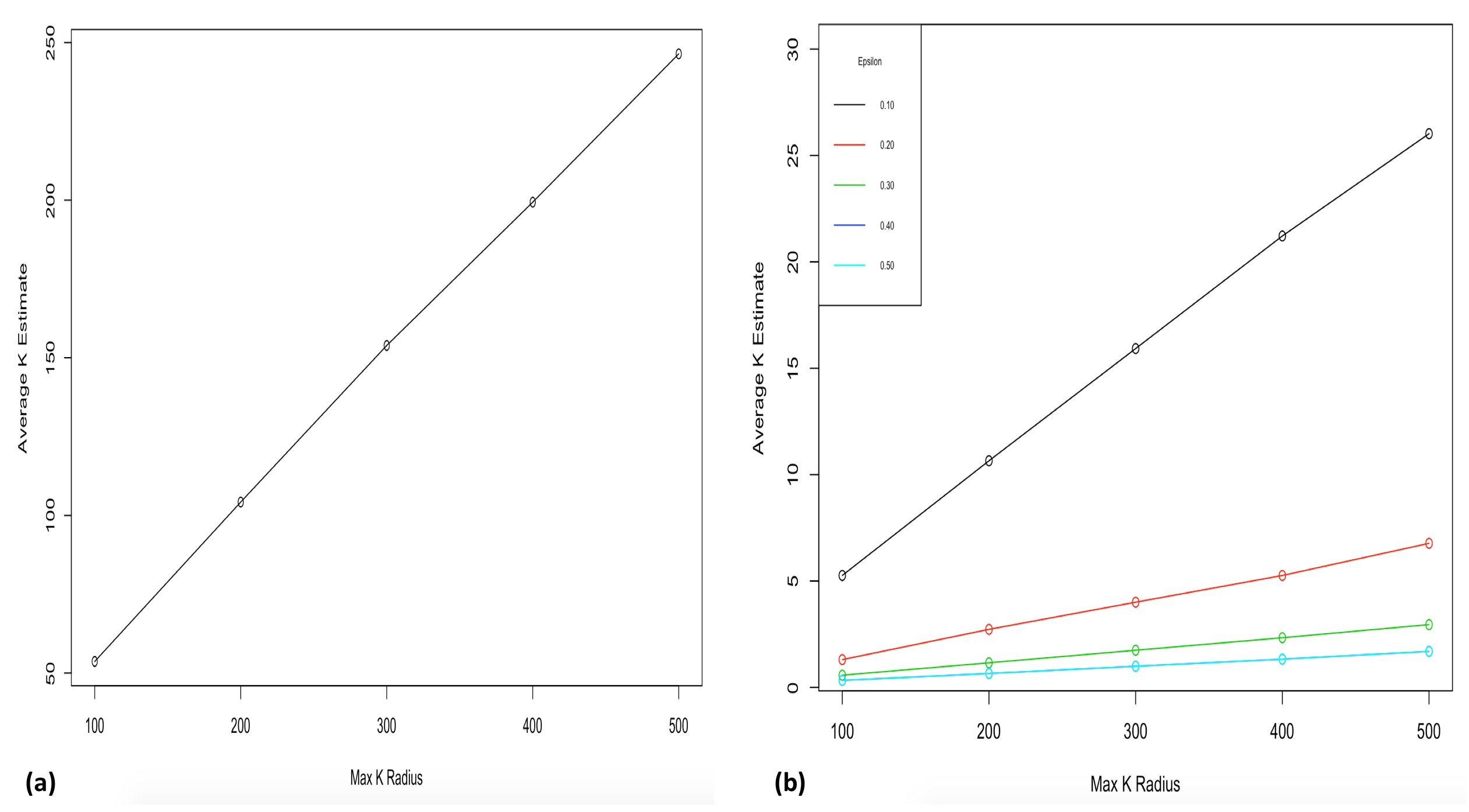}
	\caption{A plot of the Achieved K-estimate showing average k achieved vs average radius at Max K (see Section \ref{methodology}.). The diagram shows (a) Achieved K for the Donut method and (b) Achieved K for the Geo-I method.}
	\label{fig:kestimate}
\end{figure}

The Donut approach using an inner ring radius, R1 as shown in Figure \ref{fig:Donut}, sets the minimum distance of perturbation to prevent perturbation of too small distances. As can be seen this has the effect of ensuring relatively high Achieved K measures that increases linearly with Max K. On the other hand, the Geo-I approach did not guarantee stronger privacy protection over widened protection distances in our analysis. Unlike the Donut approach where Achieved K-estimate values increase linearly with perturbation distances, the Geo-I approach provides privacy that correlates to a minimized privacy budget (i.e. epsilon). As shown in Figure \ref{fig:kestimate}, the lowest Achieved K estimates (0 to 5) was recorded for epsilons at 0.3 to 0.5 whereas a steep rise to 25 was observed for epsilon at 0.1 which suggest an improved privacy with smaller privacy budgets.

\begin{figure}[h]
	\centering
	\includegraphics[width=1.0\textwidth]{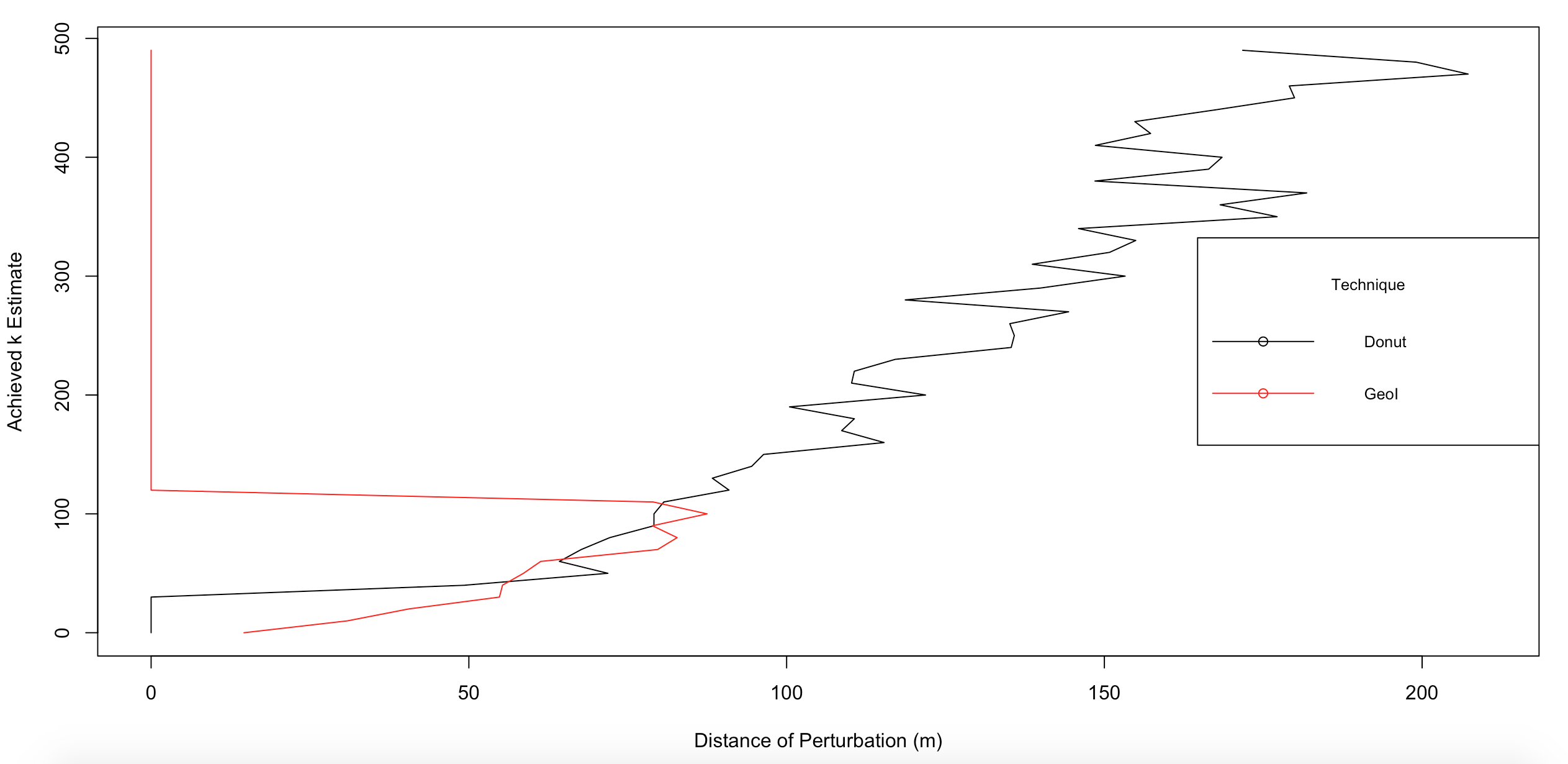}
	\caption{A plot of Average Achieved K against perturbation distances.}
	\label{fig:kestdist}
\end{figure}

Our continued analysis incrementing protection distances (with larger Max K) for Geo-I did not impact on Achieved K-estimates whereas the Donut method steadily increased with distances. Our analysis ended with a minimum privacy budget of 0.1 which performed the best of the chosen privacy budget values. Lowering the privacy budget improves privacy protection \citep{primault2014differentially, oya2017geo}. The absence of an inner ring in the Geo-I, however, allowed for the negligible distances of perturbation resulting in  weak levels of privacy. As such, when considering all levels of Max K and epsilon, the Donut method performed better in the delivery of privacy protection over the Geo-I. 

\subsubsection{Average K-estimate and Perturbation Distance}
We use the optimal performing parameters for Donut at Max K of 500 and Geo-I with Max K of 500 and a privacy budget of 0.1, to evaluate the correlation of perturbation distances and Achieved K-estimates. We break Achieved K-estimates into bins of 10 and aggregate the mean perturbation distances with both techniques. Geo-I records its maximum Achieved K-estimate at 112 and its highest perturbation distance at 87 meters as shown in Figure \ref{fig:kestdist}. Notice that despite choosing a K-Max value of 500, the Geo-I technique never produced Achieved K-estimate values near 500.
At its lowest end, an Achieved K-estimate of 0 is recorded with an average distance of about 10 meters. This denotes very weak privacy protection as negligible distances are observed with an Achieved K-estimate of less than 10, i.e. the locations are highly distinguishable. The Donut method on the other had never produces Achieved K-estimates of less than 10 and results in perturbation distances from  60 to 200 meters. In other words, the Donut does a much better job of ensuring privacy.

\subsection{Analysis of Perturbed Distance Distribution}
As mentioned before, each perturbation technique seeks to protect a respondent`s location by transposing to a random distance away from its original location. We studied the distribution of perturbed distances to evaluate whether these distances conformed to known probability distributions. If they do, then the original locations can be inferred with knowledge only of the probability distribution and its parameters (see \cite{farooq2013simulation} for details). 

\begin{figure}[h]
	\begin{subfigure}{.5\textwidth}
		\centering
		\includegraphics[width=1.05\linewidth]{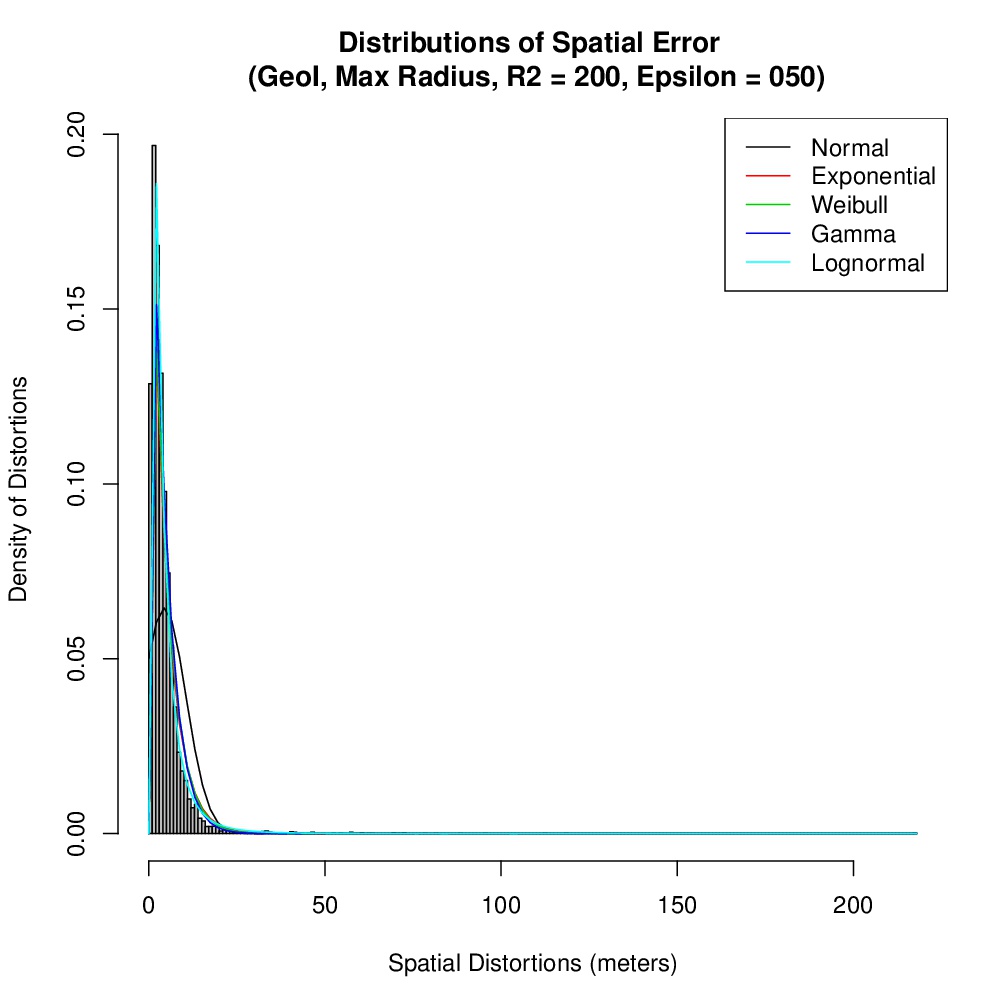}
		\caption{Geo-I}
		\label{fig:distgeoi}
	\end{subfigure}%
	\begin{subfigure}{.5\textwidth}
		\centering
		\includegraphics[width=1.05\linewidth]{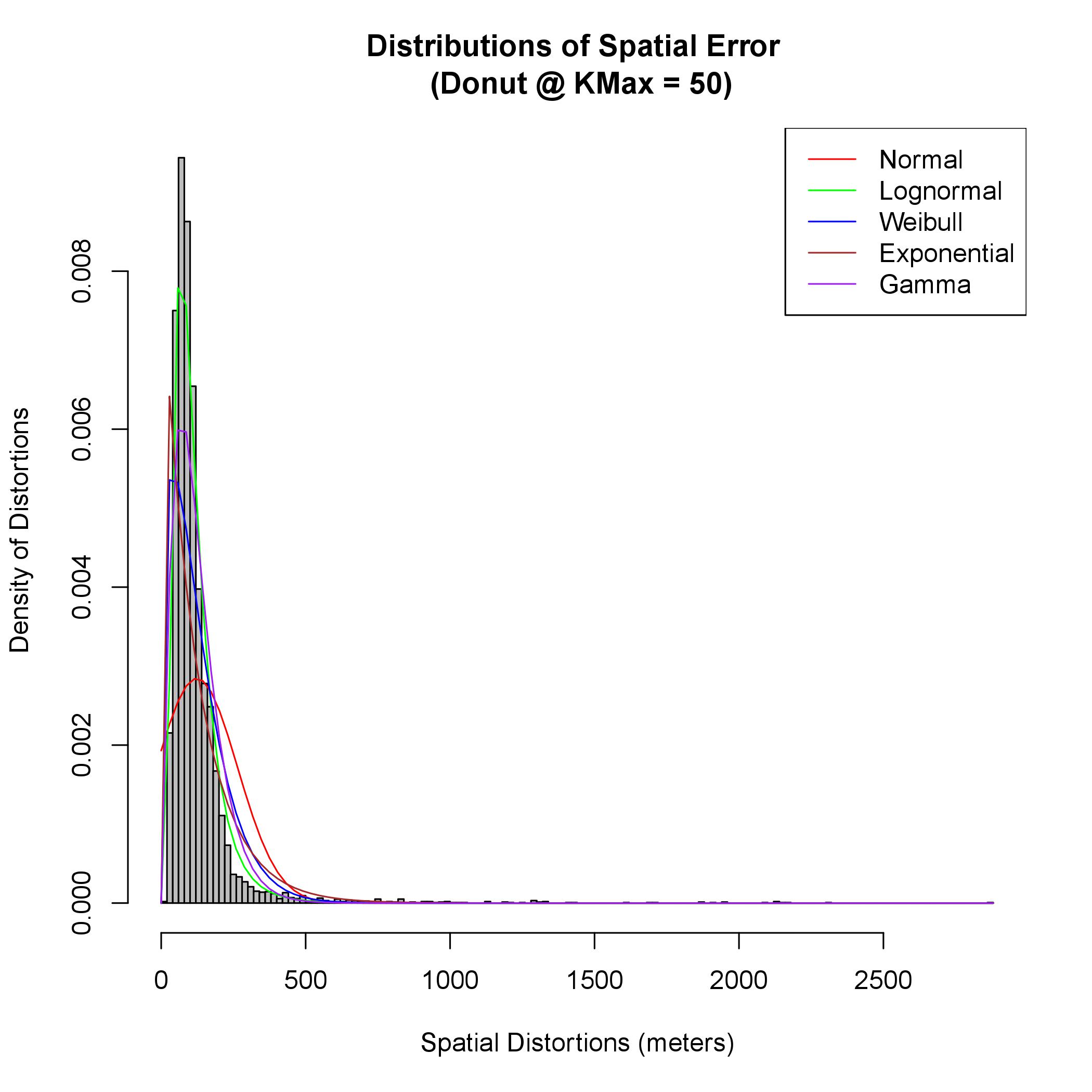}
		\caption{Donut}
		\label{fig:distDonut}
	\end{subfigure}%
	\caption{The plot of density distributions of spatial distortions for both methods.}
	\label{fig:likelihood}
\end{figure}
\begin{figure}[!h]
	\includegraphics[width=1.05\textwidth]{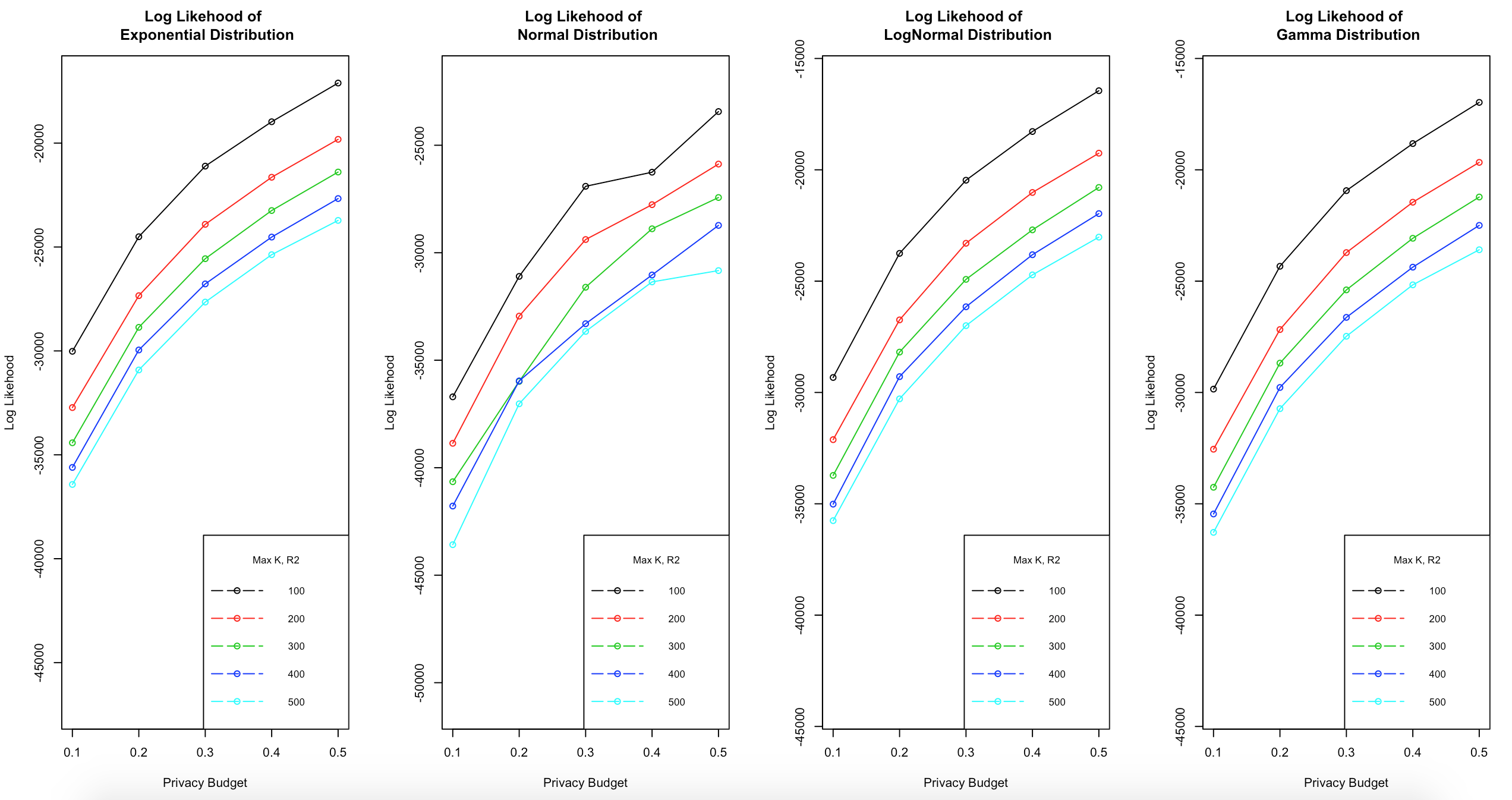}
	\caption{Log Likelihood plots for continuous distributions on Geo-I }
	\label{fig:lle_geoi}
\end{figure}
\begin{figure}[h]
	\centering
	\includegraphics[width=1.0\textwidth]{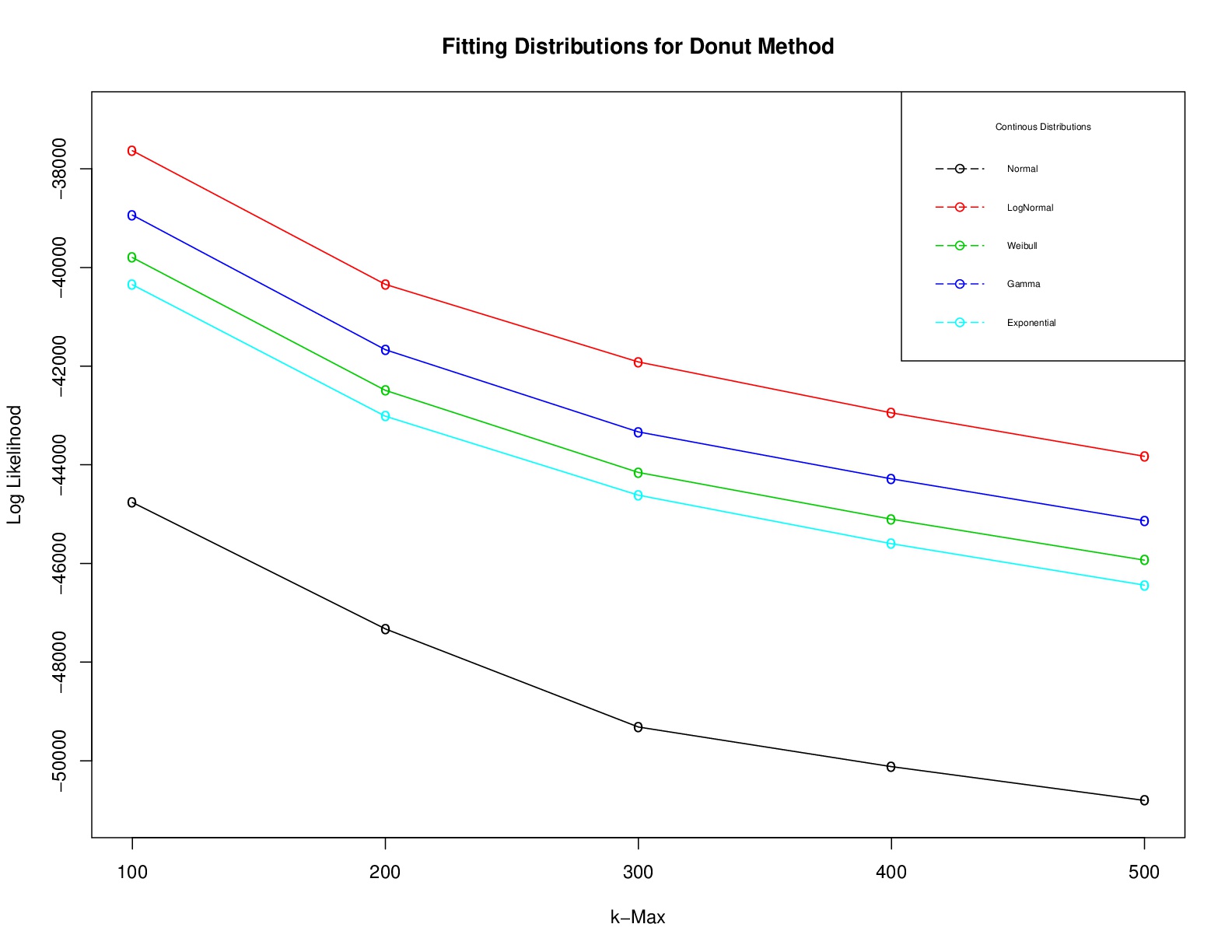}
	\caption{Log Likelihood plots for continuous distributions on Donut }
	\label{fig:lle_Donut}
\end{figure}

To do this, we first calculated the euclidean distances between the original location and its generated obfuscated location for each perturbation method. In order to evaluate whether the distributions conformed to known distributions, we fit common continuous distributions (i.e. Normal, Lognormal, Weibull, Gamma, Exponential) using maximum likelihood estimation to the perturbed distance distributions. We also recorded the maximum log-likelihood values for each of the distributions for different values of K-Max and epsilon for distances achieved by both the Donut and Geo-I methods. The lognormal distribution recorded the highest maximum log likehood values for both techniques. The distribution has its greatest density centered about the mode value, where the mode value represents a positive linear skew. We illustrate the empirical anonymization distributions of spatial distortions observed for both perturbation methods as shown in Figures \ref{fig:lle_geoi} and \ref{fig:lle_Donut}. The distribution of randomness in the spatial error fits better in a lognormal distribution as shown in the plots.  

In order to attest that the randomness of perturbation distances generated by the two approaches are closest to a lognormal distribution, we report the average maximum log likelihood statistics of the distributions of fit to the anonymized data sets for both techniques. 

\begin{table}[!h]
	\caption{A table showing average maximum likelihood values of continuous distributions fitted on anonymized data for perturbation methods.}
	\begin{center}
		\begin{tabular}{llllll}
			\hline
			\textbf{Method} & \textbf{Normal} & \textbf{Exponential} & \textbf{Weibull} & \textbf{Gamma} & \textbf{Lognormal} \\ \hline
			\hline
			Donut           & -48464.978      & -44003.698           & -43495.976       & -42671.105     & -41333.105           \\
			Geo-I            & -32442.575      & -26170.354           & -26144.260       & -26002.949     & -25521.902           \\
			\hline
		\end{tabular}
	\end{center}
	\label{table:likelihood}
\end{table}

\subsection{Analysis of Data Utility}
\begin{figure}[!h]
	\centering
	\includegraphics[width=1.0\textwidth]{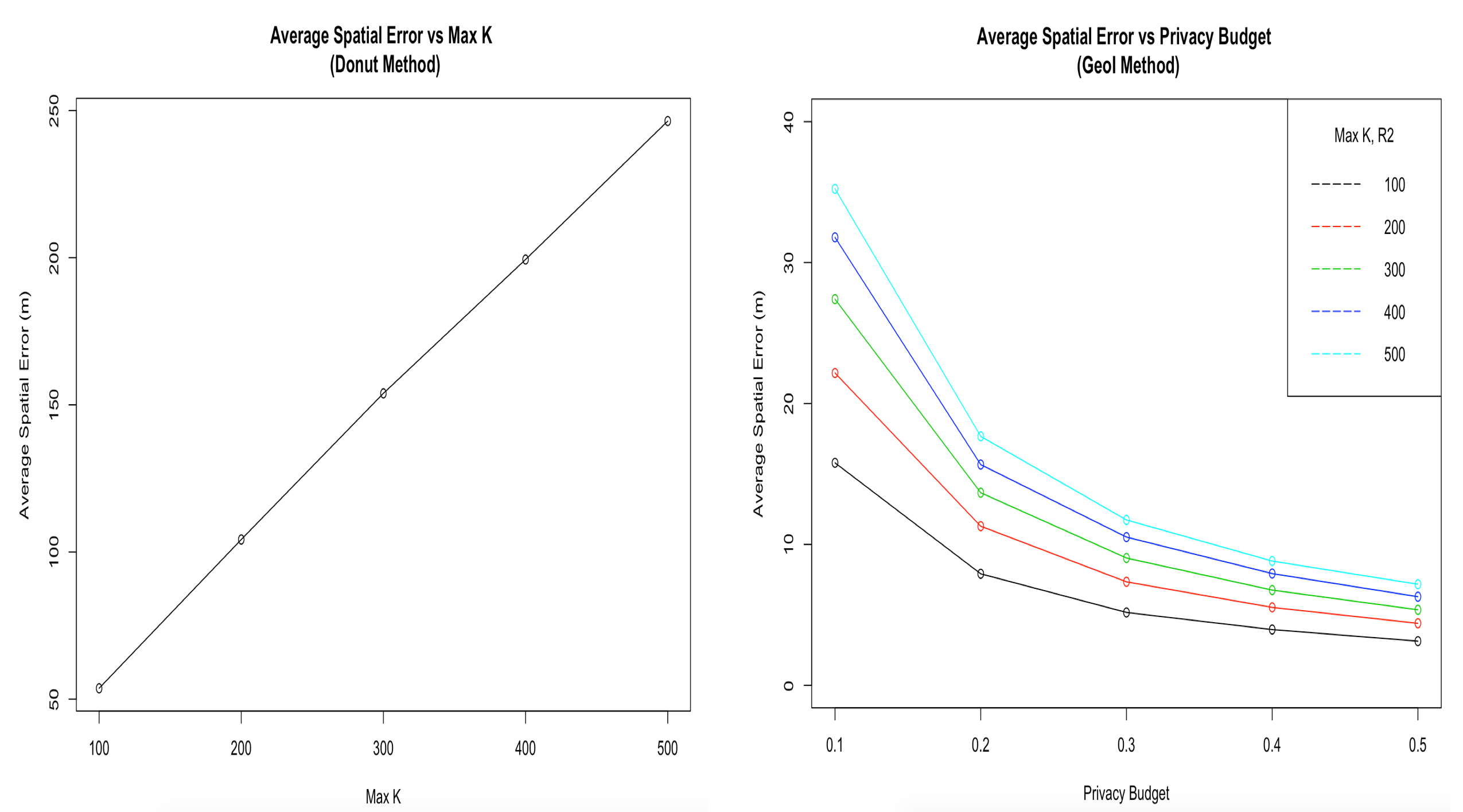}
	\caption{Spatial Error observed for Donut (left graph) and Geo-I method (right graph)}
	\label{fig:spatialerror}
\end{figure}

Finally, we calculate the average spatial error for sets of perturbation distances achieved by varying K Max (i.e. 100, 200, 300, 400, 500) and epsilon. As shown in Figure \ref{fig:spatialerror}, the Donut method shows a steady increase in the spatial error with an increase in Max-K. The magnitude of spatial error degrades the utility of the anonymized data, however. This implies that to ensure a high utility of anonymity for the Donut approach, the protection distance should be reduced so as to ensure privacy protection, while maintaining the utility of the perturbed data.

The Geo-I method exhibits a high utility on anonymized data as average spatial error decreases gradually with an increase in the privacy budget as shown in Figure \ref{fig:spatialerror}. A high average spatial error was observed at an estimate of 35 meters for the smallest value of epsilon (i.e. 0.1) to a mean protection distance of K Max at 500. Meanwhile, at a high epsilon of 0.5 applying the mean of the same K Max values, an estimate of average spatial error of only about 10 meters was observed. With this observation, the Geo-I provides a high utility relative to the Donut approach, but clearly limited privacy protection.

\section{Concluding Remarks}

In this paper, we have evaluated two popular geographic perturbation techniques (Geo-indistinguishability and Donut Geomask) to anonymize residential location data from a large-scale smartphone travel survey. Our results showed that the Donut method performs better for anonymizing location data than the Geo-I method. The degree of privacy resulting from the Donut method increased linearly with an increase of the protection distance thereby making the method sensitive to desired K-anonymity levels. The inner radius used for the Donut method that determines a minimum distance of perturbation provides a great improvement to output perturbation. As observed in our analysis, the inner radius prevented negligible perturbed distances thus this technique guarantees a strong privacy protection of sensitive locations.

On the other hand, Geo-I, which has drawn a lot of attention recently, showed much worse privacy, but did however show promise for preserving the utility of the data. Unlike the Donut method, Geo-I is not sensitive to increased protection distance, but rather to lower privacy budget (i.e. epsilon) values. This was evident in our analysis, where we experimented with a range of epsilon values. At the smallest epsilon value of 0.1, we achieved the best perturbation distance for the Geo-I at an average of 25 meters. As explained by Oya et al. \citep{oya2017geo}, the poor performance of Geo-I is attributed to the fact that counting queries in differential privacy has low sensitivity. This means that an addition or removal of a single record does not significantly affect the outcome of a query thereby ensuring high privacy achievement without introducing much noise. On the contrary, queries implemented in Geo-I demonstrate high sensitivity and therefore require large noise to achieve a high level of privacy.

Notwithstanding, while analyzing the distribution of perturbed distances, we observed that the distances of perturbation closely resembled lognormal distributions for both approaches. This means that true randomness in the resulting displacements from perturbation is not achieved. With this in mind, an adversary with prior knowledge of the perturbation distance distributions would be able to reversely identify a real location from its perturbed point within a high degree of precision. 

We find interest in advancing our research into improving the protection efficiency of Geo-I by introducing the concept of inner radius as implemented in the Donut method. A potential solution to improve Geo-I privacy performance which we seek to investigate, is to design its location queries to have lower sensitivity. Further investigation of Geo-I using smaller protection budgets might also prove fruitful. Also, for the Donut method, we would like to work on incorporating location semantics in determining the optimal radius of perturbation as an addition to the existing computation by population density.

We also acknowledge the scope of this work is focused on the anonymization of  independent point samples and that there is also a potential interest for protecting privacy in the context of multi-point data such as trajectories. We hope to further our research into location-privacy protection mechanisms that address multiple trips points and trajectories of mobile users in the future.

\section{Acknowledgements}

This research has been funded by the Social Sciences and Humanities Research Council of Canada (SSHRC).

\clearpage


\section{References}
\bibliographystyle{plainnat}
\bibliography{references.bib}

%
%
%
%

\end{document}